\documentclass[12pt]{article}
\usepackage{graphicx}

\def\vub{IIHE - VUB\\
Pleinlaan 2, 1050 Brussel, Belgie}

\def\Title#1{\begin{center} {\Large #1 } \end{center}}
\def\Author#1{\begin{center}{ \sc #1} \end{center}}
\def\Address#1{\begin{center}{ \it #1} \end{center}}

\newenvironment{Abstract}{\begin{quotation}  }{\end{quotation}}
\newenvironment{Presented}{\begin{quotation} \begin{center} 
             PRESENTED AT\end{center}\bigskip 
      \begin{center}\begin{large}}{\end{large}\end{center} \end{quotation}}

\def\beq{\begin{equation}}
\def\eeq#1{\label{#1}\end{equation}}
\def\eeqn{\end{equation}}

\def\beqa{\begin{eqnarray}}
\def\eeqa#1{\label{#1}\end{eqnarray}}
\def\eeqan{\end{eqnarray}}

\let\bar=\overbar

\def\Dslash{\not{\hbox{\kern-4pt $D$}}}
\def\dslash{\not{\hbox{\kern-2pt $\del$}}}

\def\msb{{\bar{\ssstyle M \kern -1pt S}}}

\begin{document}
\begin{titlepage}

\vfill
\Title{Search for flavour-changing neutral currents with top quarks}
\vfill
\Author{Kirill Skovpen\\(on behalf of the ATLAS and CMS Collaborations)}
\Address{\vub}
\vfill
\begin{Abstract}
Flavour-changing neutral currents are extremely rare processes in the
standard model that can be sensitive to various new physics effects.
A summary of the latest experimental results from the LHC
experiments is given. Preliminary results of sensitivity studies 
for future colliders are also discussed.
\end{Abstract}
\vfill
\begin{Presented}
$10^{th}$ International Workshop on Top Quark Physics TOP2017\\
Braga, Portugal, September 17--22, 2017
\end{Presented}
\vfill
\end{titlepage}
\def\thefootnote{\fnsymbol{footnote}}
\setcounter{footnote}{0}

\section{Introduction}

In the SM, flavor-changing neutral currents (FCNC) are forbidden
at tree level and are strongly suppressed in higher orders
by the Glashow-Iliopoulos-Maiani (GIM) mechanism~\cite{GIM}. 
The predicted branching fractions for top quark FCNC decays are
expected to be $\mathcal{O}(10^{-12}$--$10^{-16})$~\cite{FCNC1,FCNC2,FCNC3}.
In various extensions of the SM the FCNC effects are significantly
enhanced and can be directly probed at the LHC~\cite{FCNC3,FCNC4,FCNC5}.
Top quark FCNC interactions can be written with the following
lagrangian:

\begin{eqnarray}
\mathcal{L} =
\sum_{q=u,c}\big[\sqrt{2}g_{s}\frac{\kappa_{gqt}}{\Lambda}\bar{t}\sigma^{\mu\nu}T_{a}
(f^{L}_{Gq}P_{L}+f^{R}_{Gq}P_{R})qG_{\mu\nu}^{a}+ \nonumber \\
+\frac{g}{\sqrt{2}c_{W}}\frac{\kappa_{zqt}}{\Lambda}\bar{t}\sigma^{\mu\nu}
(f^{L}_{Zq}P_{L}+f^{R}_{Zq}P_{R})qZ_{\mu\nu}+
\frac{g}{4c_{W}}\zeta_{zqt}\bar{t}\gamma^{\mu}
(f^{L}_{Zq}P_{L}+f^{R}_{Zq}P_{R})qZ_{\mu}- \nonumber \\
-e\frac{\kappa_{\gamma qt}}{\Lambda}\bar{t}\sigma^{\mu\nu}
(f^{L}_{\gamma q}P_{L}+f^{R}_{\gamma q}P_{R})qA_{\mu\nu}+ \nonumber \\
+\frac{g}{\sqrt{2}}\bar{t}\kappa_{Hqt}
(f^{L}_{Hq}P_{L}+f^{R}_{Hq}P_{R})qH
\big]+h.c.,
\end{eqnarray}

\noindent where $P_{L}$ and $P_{R}$ are chirality projectors in spin space,
$\kappa_{Xqt}$ is the effective coupling
for $tXq$ vertex ($X=g,Z,\gamma,H$), $\zeta_{Zqt}$ is the additional
effective coupling for $tZq$ vertex, $f_{Xq}^L$ and $f_{Xq}^R$ are left and right-handed complex
chiral parameters with a unitarity constraint of $|f^L_{Xq}|^2 +
|f^R_{Xq}|^2 = 1$. The ATLAS~\cite{ATLASDET} and CMS~\cite{CMSDET} experiments at the LHC
study top FCNC anomalous interactions in the decays of top quarks in top quark pair
production, as well as in singly produced top quarks.

\section{Overview of results}

The $tgq$ anomalous interactions are studied by CMS~\cite{CMStgq} and
ATLAS~\cite{ATLAStgq} in single top quark events. The event signature
considered by CMS in the analysis of data collected at 7 and 8 TeV includes the
requirement of one isolated lepton, the presence of a significant amount of transverse
missing energy ($E^{miss}_{T}$), exactly one b and one non-b jet.
The dominant background is the $t\bar{t}$+jets
production. A Bayesian neural network
technique is used to separate signal from background events.
The observed~(expected) 95\% confidence level (CL) upper limits are
found to be $\mathcal{B}(t \to gu) < 2.0~(2.8) \times 10^{-5}$ and
$\mathcal{B}(t \to gc) < 4.1~(2.8) \times 10^{-4}$.
Selection criteria used in the analysis by ATLAS at 8 TeV include a veto on
additional reconstructed jets with requiring only one b jet to be
present in the final state. The background is dominated by $W$+jets
production.
Separation between signal and background
events is achieved with a neural network classifier. The resultant
observed~(expected) limits are $\mathcal{B}(t \to gu) < 4.0~(3.5) \times 10^{-5}$ 
and $\mathcal{B}(t \to gc) < 2.0~(1.8) \times 10^{-4}$.

The $t \rightarrow Z q$ FCNC decays are probed by ATLAS in top quark
pair events at 13 TeV~\cite{ATLAStZq}. The event topology includes 
the presence of three isolated leptons, $E^{miss}_{T} > 40$~GeV, exactly
one b jet and at least one non-b jet. Several control regions are
defined for each of the dominant background processes: $WZ$+jets,
$ZZ$+jets, $ttZ$, and non-prompt leptons. Simultaneous maximum likelihood fit over control
and signal regions is performed to extract the exclusion limits,
$\mathcal{B}(t \to Zu) < 1.7~(2.4) \times 10^{-4}$ and $\mathcal{B}(t
\to Zc) < 2.3~(3.2) \times 10^{-4}$. The analysis done by CMS with
8 TeV data explores a similar final state and additionally includes
a single top associated FCNC production with a Z boson in the simulation of
signal events~\cite{CMStZq}. A boosted decision tree (BDT) discriminant is defined to
suppress background events. The resultant limits are $\mathcal{B}(t \to
Zu) < 2.2~(2.7) \times 10^{-4}$ and $\mathcal{B}(t
\to Zc) < 4.9~(11.8) \times 10^{-4}$.

The $t\gamma q$ anomalous interactions are probed by CMS at 8 TeV in
events with single top quarks produced in association with a photon~\cite{CMStgammaq}.
Event selection criteria includes the presence of one isolated lepton,
one isolated photon, $E^{miss}_{T} > 30$~GeV, and up to one b jet. The
dominant $W\gamma$ background along with $W$+jets events are
suppressed with a BDT discriminant. The resultant exclusion limits are
$\mathcal{B}(t \to \gamma u) < 1.3~(1.9) \times 10^{-4}$ and
$\mathcal{B}(t \to \gamma c) < 2.0~(1.7) \times 10^{-3}$.

The $tHq$ FCNC processes are studied by ATLAS in top quark
pair events with $t \rightarrow qH, H \rightarrow \gamma\gamma$ at 13
TeV~\cite{TopHiggsATLAS13TeV}. The analysis explores the final state with two isolated
photons. For leptonic top quark decays the selection criteria includes
the requirement of one isolated lepton, exactly one b jet, and at least
one non-b jet. In case of hadronic top quark decays the analysis
selects events with no isolated leptons, at least one b jet, and at
least three additional non-b jets. The dominant background processes
are associated with the production of non-resonant $\gamma\gamma$+jets, $t\bar{t}$+jets and
$W$+$\gamma\gamma$ events. The analysis is done in the
diphoton mass window of $100 < m_{\gamma\gamma} < 160$~GeV.
The resultant limits are found to be $\mathcal{B}(t \to Hu) < 2.4~(1.7) \times
10^{-3}$ and $\mathcal{B}(t \to Hc) < 2.2~(1.6) \times 10^{-3}$.
The $tHq$ anomalous couplings are probed by CMS in $H \rightarrow
b\bar{b}$ channel in top quark pair events, as well as in single top
associated production with a Higgs boson, at 13
TeV~\cite{TopHiggsCMS13TeV}. The event selection includes the
requirement of one isolated lepton, at least two b jets, and at least
one additional non-b jet. The dominant $t\bar{t}$ background is
suppressed with a BDT discriminant to set the exclusion limits of
$\mathcal{B}(t \to Hu) < 4.7~(3.4) \times 10^{-3}$ and $\mathcal{B}(t
\to Hc) < 4.7~(4.4) \times 10^{-3}$. The combination of results
obtined by CMS at 8 TeV results in $\mathcal{B}(t \to Hu) < 5.5~(4.0) \times 10^{-3}$ and
$\mathcal{B}(t \to Hc) < 4.0~(4.3) \times 10^{-3}$~\cite{CMS8}, while a similar
combination of ATLAS results yields $\mathcal{B}(t \to Hu) < 4.5~(2.9)
\times 10^{-3}$ and $\mathcal{B}(t \to Hc) < 4.6~(2.5) \times
10^{-3}$~\cite{ATLAS8}.

The upcoming upgrade of the LHC, the High-Luminosity LHC (HL-LHC)
project, is expected to introduce a substantial increase in the peak
luminosity with accumulating a larger data set to reach
$\simeq~3~ab^{-1}$ by the end of the data taking. Preliminary
sensitivity studies for the HL-LHC suggest the expected 95\% CL upper limit of
$\mathcal{B}(t \to Zq) < 2.4 \times 10^{-5}$~\cite{HLLHCtZq}.
Similar studies for $t\gamma q$ and $tHq$ yield $\mathcal{B}(t \to \gamma u) <
2.7 \times 10^{-5}$, $\mathcal{B}(t \to \gamma c) < 2.0 \times 10^{-4}$~\cite{HLLHCtgammaq},
and $\mathcal{B}(t \to Hq) < \mathcal{O}(10^{-4})$~\cite{HLLHCtZq,HLLHCtHq}.
A proposed experiment with the focus on the study of deep inelastic lepton-hadron scattering, 
a Large Hadron electron Collider (LHeC), is expected to
provide a competitive sensitivity to $t\gamma q$ and $tZq$ searches
with the expected limits of $\mathcal{B}(t \to \gamma u) < 1.62 \times 10^{-5}$
and $\mathcal{B}(t \to \gamma c) < 1.15 \times 10^{-4}$~\cite{LHeCtgammaq2017}.
The $tHq$ projections for LHeC yield $\mathcal{B}(t \to Hu) < 7.3 \times 10^{-4}$~\cite{LHeCtHq}.
Future Circular Collider for electron-positron (FCC-ee) and proton-proton (FCC-hh) collisions is a planned set
of research activities targeted at the timescale beyond the operation time of the HL-LHC.
Preliminary projections suggest the following expected limits: $\mathcal{B}(t \to
Zq) < 5.6 \times 10^{-6}$ (FCC-ee)~\cite{FCCeetZq}, $\mathcal{B}(t \to
Zq) < \mathcal{O}(10^{-7})$ (FCC-hh)~\cite{FCChh}, $\mathcal{B}(t \to
\gamma q) < 3.6 \times 10^{-6}$ (FCC-ee)~\cite{FCCeetZq}, and
$\mathcal{B}(t \to \gamma q) < \mathcal{O}(10^{-7})$ (FCC-hh)~\cite{FCChh}.
Future linear electron-positron colliders (ILC/CLIC) can also provide a
competitive sensitivity to top FCNC studies with the expected limits of $\mathcal{B}(t \to Zq) < \mathcal{O}(10^{-5})$~\cite{ILCtZq},
$\mathcal{B}(t \to \gamma q) < \mathcal{O}(10^{-5})$~\cite{ILCtZq}, and
$\mathcal{B}(t \to Hq) < \mathcal{O}(10^{-5})$~\cite{ILCtHq}.

\section{Conclusion}

\begin{figure}[hbtp]
  \begin{center}
  \includegraphics[width=0.50\linewidth]{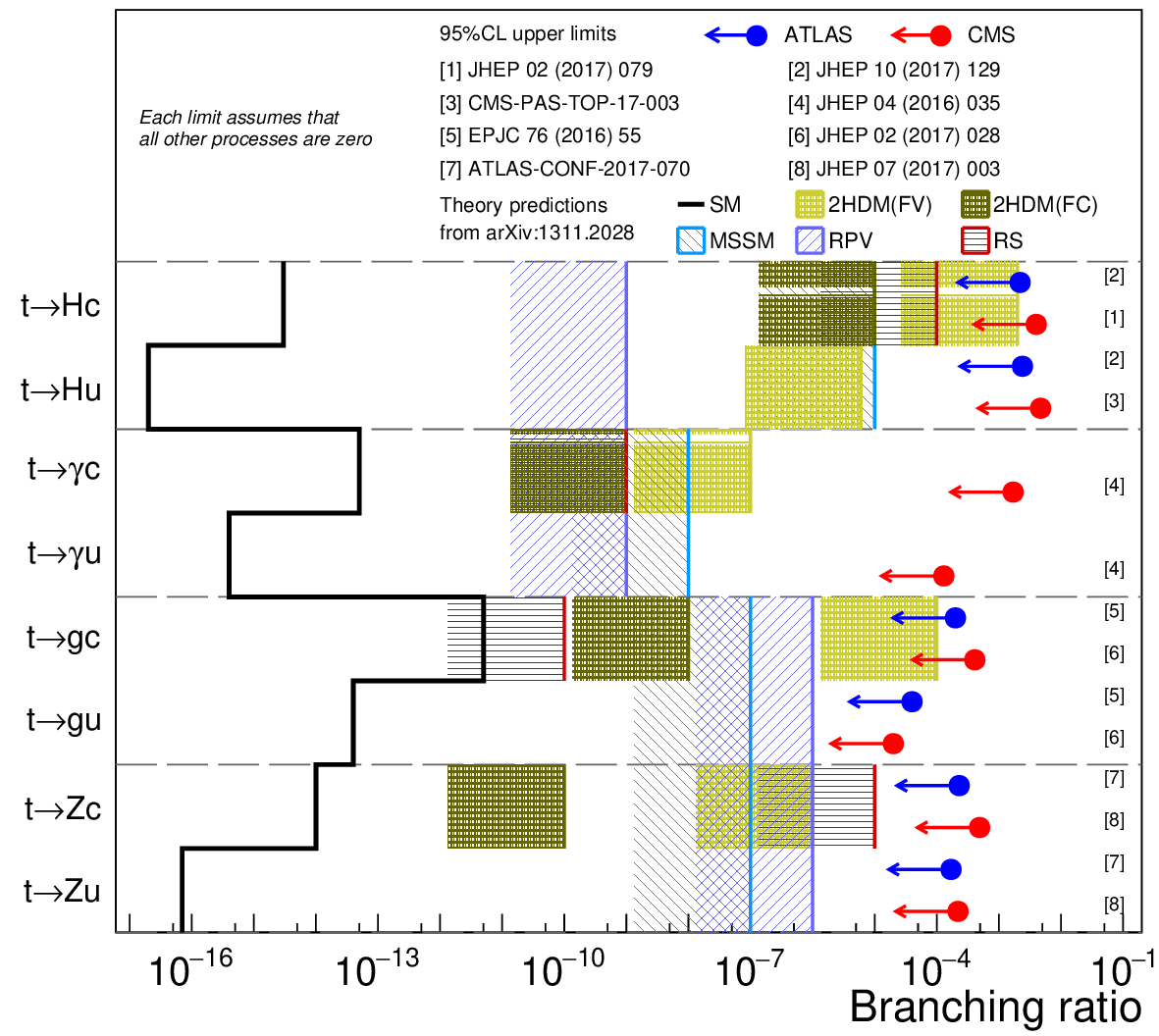}
  \caption{Summary of branching fraction limits for top quark FCNC
  decays, compared to the SM and new physics models predictions.
  Experimental results are shown
  from~\cite{CMStgq,ATLAStgq}($tgq$),~\cite{ATLAStZq,CMStZq}($tZq$),~\cite{CMStgammaq}($t\gamma
  q$), and~\cite{TopHiggsATLAS13TeV,TopHiggsCMS13TeV,CMS8}($tHq$).}
  \label{fig:summary}
  \end{center}
\end{figure}

\begin{figure}[hbtp]
  \begin{center}
  \includegraphics[width=0.50\linewidth]{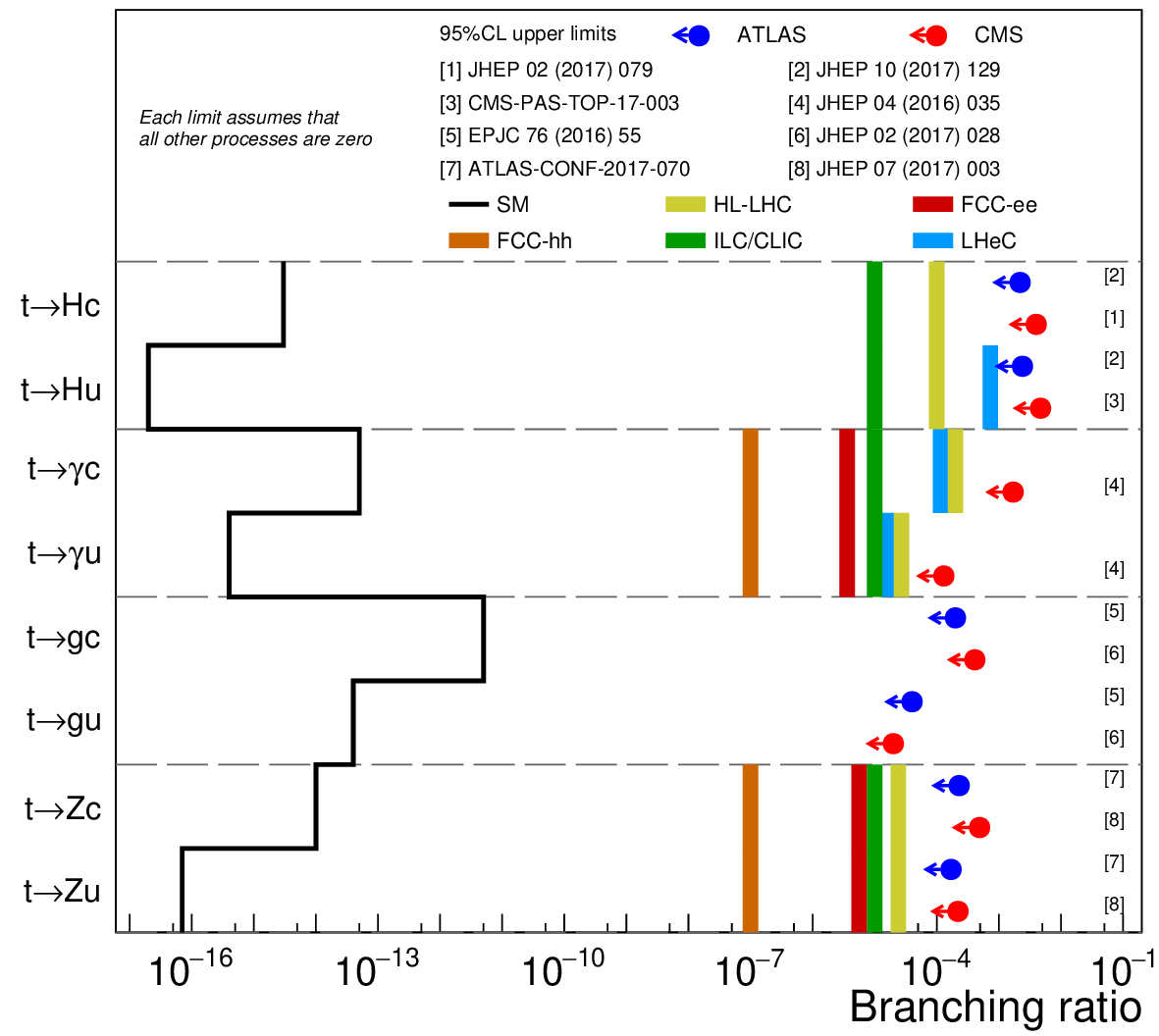}
  \caption{Summary of branching fraction limits for top quark FCNC
  decays, compared to the SM predictions, as well as to various
  projections for future experiments.   Experimental results are shown
  from~\cite{CMStgq,ATLAStgq}($tgq$),~\cite{ATLAStZq,CMStZq}($tZq$),~\cite{CMStgammaq}($t\gamma
  q$), and~\cite{TopHiggsATLAS13TeV,TopHiggsCMS13TeV,CMS8}($tHq$).
  Projections are taken from~\cite{HLLHCtZq,HLLHCtgammaq,HLLHCtHq}(HL-LHC),~\cite{LHeCtgammaq2017,LHeCtHq}(LHeC),~\cite{FCChh,FCCeetZq}(FCC),
  and~\cite{ILCtZq}(ILC).}
  \label{fig:future}
  \end{center}
\end{figure}

The most stringent limits on top quark FCNC branching fractions
obtained at the LHC are compared to theoretical predictions in the
SM, as well as to various new physics models. A summary of these
results is presented in Figure~\ref{fig:summary}, and several projections
of the upper limits for future experiments are summarized in
Figure~\ref{fig:future}. A large number of experimental results
on top quark FCNC interactions is available from the LHC. The latest results
obtained at 13 TeV signficantly improve the upper limits set with the 8 TeV
data. The preliminary results of future projections represent good 
prospects for pushing top FCNC boundaries to even higher constraints.

\begin{thebibliography}{99}

\bibitem{GIM}
S. L. Glashow, J. Iliopoulos and L. Maiani, Phys. Rev. D {\bf 2}
(1970) 1285.

\bibitem{FCNC1}
G. Eilam, J. L. Hewett, and A. Soni, Phys. Rev. D {\bf 44}
(1991) 1473.

\bibitem{FCNC2}
B. Mele, S. Petrarca, and A. Soddu, Phys. Lett. B {\bf 435}
(1998) 401.

\bibitem{FCNC3}
J. A. Aguilar-Saavedra, Acta Phys. Polon. B {\bf 35}
(2004) 2695, hep-ph/0409342.

\bibitem{FCNC4}
F. Larios, R. Martinez, M. A. Perez, Int. J. Mod. Phys. A {\bf 21}
(2006) 3473, hep-ph/0605003.

\bibitem{FCNC5}
K. Agashe et al., hep-ph/1311.2028.

\bibitem{ATLASDET}
ATLAS Collaboration, JINST 3 S08003 (2008).

\bibitem{CMSDET}
CMS Collaboration, JINST 3 S08004 (2008).

\bibitem{CMStgq}
CMS Collaboration, JHEP {\bf 02} (2017) 028, hep-ex/1610.03545.

\bibitem{ATLAStgq}
ATLAS Collaboration, Eur. Phys. J. C {\bf 76} (2016) 55, hep-ex/1509.00294.

\bibitem{ATLAStZq}
ATLAS Collaboration, ATLAS-CONF-2017-070,\\
http://cds.cern.ch/record/2285808.

\bibitem{CMStZq}
CMS Collaboration, JHEP {\bf 07} (2017) 003, hep-ex/1702.01404.

\bibitem{CMStgammaq}
CMS Collaboration, JHEP {\bf 04} (2016) 035, hep-ex/1511.03951.

\bibitem{TopHiggsATLAS13TeV}
ATLAS Collaboration, JHEP {\bf 10} (2017) 129, hep-ex/1707.01404.

\bibitem{TopHiggsCMS13TeV}
CMS Collaboration, CMS-PAS-TOP-17-003,\\
http://cds.cern.ch/record/2284743.

\bibitem{CMS8}
CMS Collaboration, JHEP {\bf 02} (2017) 079, hep-ex/1610.04857.

\bibitem{ATLAS8}
ATLAS Collaboration, JHEP {\bf 12} (2015) 061, hep-ex/1509.06047.



\bibitem{HLLHCtZq}
ATLAS Collaboration, ATL-PHYS-PUB-2016-019,\\
http://cds.cern.ch/record/2209126.

\bibitem{HLLHCtgammaq}
CMS Collaboration, CMS-PAS-FTR-16-006,\\
http://cds.cern.ch/record/2262606.

\bibitem{HLLHCtHq}
ATLAS Collaboration, ATL-PHYS-PUB-2013-012,\\
http://cds.cern.ch/record/1604506.



\bibitem{LHeCtgammaq2017}
I. T. Cakir et al., hep-ph/1705.05419.

\bibitem{LHeCtHq}
X. Wang, H. Sun, X. Luo, Adv. High Energy Phys. {\bf 4693213}
(2017), hep-ph/1703.02691.


\bibitem{FCChh}
M. L. Mangano et al., hep-ph/1607.01831.

\bibitem{FCCeetZq}
H. Khanpour, S. Khatibi, M. K. Yanehsari, M. M. Najafabadi, hep-ph/1408.2090.


\bibitem{ILCtZq}
G. Moortgat-Pick et al., Phys. Rept. {\bf 460} (2008) 131, hep-ph/0507011.

\bibitem{ILCtHq}
A. \v{Z}arnecki (on behalf of the CLICdp collaboration and the ILC
Physics and Detector Study), PoS(ICHEP2016) (2016) 666, hep-ex/1611.04492.

\end{thebibliography}
\end{document}